\documentclass[11pt,a4paper]{scrartcl}

\usepackage{CLICdp}
\usepackage{float}
\usepackage{amsmath}

\usepackage{CLICdp_definitions}

\usepackage{xpatch}
\usepackage{makecell}
\usepackage{multicol}
\usepackage{numprint}

\makeatletter
\xpatchcmd\@HepConStyle
 {\edef\@upcode{\updefault}}
 {\ifdefined\shapedefault\edef\@upcode{\shapedefault}\else\edef\@upcode{\updefault}\fi}
 {}{}
\makeatother

\usepackage{arydshln}

\newcommand{\MeV}{\ensuremath{\text{MeV}}\xspace}
\newcommand{\GeV}{\ensuremath{\text{GeV}}\xspace}
\newcommand{\TeV}{\ensuremath{\text{TeV}}\xspace}
\newcommand{\tabt}[1]{\multicolumn{1}{c}{#1}}
\newcommand{\tabtt}[1]{\multicolumn{2}{c}{#1}}
\newcommand{\mH }{\ensuremath{m_{\PH}}\xspace}
\newcommand{\GH }{\ensuremath{\Gamma_{\PH}}\xspace}
\newcommand{\BR}{\ensuremath{BR}\xspace}

\newcommand{\gHZZ}{\ensuremath{g_{\PH\PZ\PZ}}\xspace}
\newcommand{\gHWW}{\ensuremath{g_{\PH\PW\PW}}\xspace}
\newcommand{\gHtt}{\ensuremath{g_{\PH\PQt\PQt}}\xspace}
\newcommand{\gHbb}{\ensuremath{g_{\PH\PQb\PQb}}\xspace}
\newcommand{\gHcc}{\ensuremath{g_{\PH\PQc\PQc}}\xspace}
\newcommand{\gHTauTau}{\ensuremath{g_{\PH\PGt\PGt}}\xspace}
\newcommand{\gHMuMu}{\ensuremath{g_{\PH\PGm\PGm}}\xspace}

\title{CLIC Higgs coupling prospects with 100\,Hz operation}

\clicdpnote{2025}{002}  

\date{\today}

\addauthor{A.~Robson}{\institute{1}}
\addauthor{P.~Roloff}{\institute{2}}
\addauthor{J.~de Blas}{\institute{3}}
\addinstitute{1}{University of Glasgow, Glasgow, United Kingdom}
\addinstitute{2}{CERN, Geneva, Switzerland}
\addinstitute{3}{Universidad de Granada, Granada, Spain}

\abstract{The staging scenario for CLIC has been updated
  following new studies of the beam emittance
  through the accelerator chain, which has resulted in higher expected luminosities, and a change in baseline to a 100\,Hz repetition rate at the initial energy stage.
  Here, the Higgs coupling sensitivities are updated for the new staging
  plan.}

\graphicspath{ {./logos/}{./figures/} }

\begin{document}

\titlepage

\section{Introduction}

The Compact Linear Collider, CLIC, offers high-energy \epem\ collisions up to centre-of-mass energies of 3\,TeV \cite{clic-study}.
A rich programme of Higgs and top-quark physics is uniquely provided by the initial energy
stage around $\roots=380\,\GeV$; this is supplemented at the higher-energy stages
by increased precision in Higgs and top-quark physics, and further reach to Beyond Standard Model (BSM)
effects \cite{Abramowicz:2016zbo,Abramowicz:2018rjq,deBlas:2018mhx}.

\section{Staging}
\label{sec:staging}

Since the publication of the CLIC Project Implementation Plan in 2018 \cite{ESU18PiP}, the baseline luminosity for CLIC operating at $\roots=380\,\GeV$ has been updated according to new studies \cite{ESU25readinessreport}.  These include improved simulation of the beam emittance through the accelerator chain, which has resulted in expected luminosities that are around 50\% higher, and a change in baseline repetition rate from 50\,Hz to 100\,Hz at the initial energy stage, giving another factor of two increase in luminosity.
This doubling of the repetition rate can be achieved without major design changes, and with an increase in the overall power consumption of around 60\%.
The implications of the improved accelerator design on the CLIC detector model are considered in Ref.~\cite{detectorprogress}, and conclude that the current detector model remains matched to the accelerator environment.
Another update to the CLIC baseline design is the incorporation of two beam delivery systems and interaction points, such that the total luminosity could be shared between two experiments.

The new baseline is given in \cref{table:lumi:CLIC}.  One of the advantages in a linear collider approach is its flexibility: at each stage the physics and technology landscape can be reviewed, and a decision taken either to realise a subsequent energy stage, or to proceed to a different collider technology.  
While the CLIC physics studies have been carried out for the reference energies of $\sqrt{s}=380$\,GeV, 1.5\,TeV, and 3\,TeV, Ref.~\cite{ESU25readinessreport} also contains some discussion of running at alternative energies ($\sqrt{s}=91, 250$, and $550$\,GeV).

\vspace*{2mm}
\begin{table}[hbt]
\centering{
\begin{tabular}{ l|c|c|c } 
  \toprule
  $\sqrt{s}$ [GeV] &  380  &  1500    & 3000 \\
  \midrule
  Repetition frequency [Hz]  & 100 (50)    &  50      & 50   \\
  Luminosity [$10^{34}$\,cm$^{-2}$s$^{-1}$] &
                      4.5 (2.25)           & 3.7   & 5.9  \\
  Run time [years]   & 10                   & 10    & 8  \\
  Integrated luminosity [$\abinv$] & 4.3 (2.2)& 4     & 5 \\
  \bottomrule
\end{tabular}
}
\caption{The updated baseline CLIC operation model (December 2024). Two options for 380\,GeV-running are given, with 100\,Hz and 50\,Hz repetition rates, respectively.  The new baseline is 100\,Hz running.  The integrated luminosity values correspond to 185 days of physics per year and 75\% operational efficiency, i.e. $1.2\times10^7$ seconds per year.  A ramp-up of 10\%, 30\%, and 60\% in delivered luminosity is assumed over the first three years at 380\,\GeV.  A ramp-up of 25\% and 75\% is assumed over the first two years at 1.5\,TeV.
\label{table:lumi:CLIC}}
\end{table}

\section{Higgs couplings}
\label{sec:higgs}

\subsection{Summary of Higgs observables}

Extensive studies of the CLIC sensitivities to Higgs couplings have been reported
previously in \cite{Abramowicz:2016zbo}, where details of the analyses and
the extraction of Higgs observables through combined
fitting can be found{\footnote{Note that earlier studies assumed an energy staging of $\sqrt{s}=350\,\GeV$, 1.4\,\TeV, and 3\,\TeV; those energy stages are used for the results presented here, but with results scaled to the updated integrated luminosities.}}.
Sensitivities obtained assuming the previous CLIC luminosity baseline can be found in \cite{Robson:2018zje}.

Precisions on the Higgs observables for the new staging scenario are given in \autoref{tab:GlobalFit:Input350} for
the first energy stage, and in \autoref{tab:GlobalFit:Input143} for the two
higher-energy stages.  These individual results assume unpolarised beams.
It should be noted that since these full-simulation studies were performed there have been extensive improvements in reconstruction techniques, for example in heavy-flavour tagging and tau reconstruction, the implementation of which would lead to enhanced sensitivities.

Measurement of the cross section for double-Higgs production at 1.4 and 3\,TeV 
gives sensitivity to the trilinear Higgs self-coupling $\lambda$.
This is unchanged from that reported in Ref.~\cite{Roloff:2019crr}, with an ultimate precision on 
 $\lambda$ of $[-8\%,+11\%]$.


\begin{table*}[htp]\centering
  \begin{tabular}{lllccc}\toprule
                        &                                                           &                              & \tabtt{Statistical precision} & \\ \cmidrule(l){4-5}
        \tabt{Channel}  & \tabt{Measurement}                                        & \tabt{Observable}            & $350\,\GeV$ (50\,Hz) & $350\,\GeV$ (100\,Hz)  & Reference \\ 
                        &                                                           &                              &  $2.2\,\abinv$  & $4.3\,\abinv$  & \cite{Abramowicz:2016zbo} \\ \midrule
    $\PZ\PH$            & Recoil mass distribution                                  & $\mH$                        & $52\,\MeV$   & $38\,\MeV$    & \cite{Abramowicz:2016zbo} \\
    $\PZ\PH$            & $\sigma(\PZ\PH)\times \BR(\PH\to\text{invisible})$        & $\Gamma_\text{inv}$           & $0.3\,\%$    & $0.2\,\%$     & \cite{Abramowicz:2016zbo} \\ \midrule
    $\PZ\PH$            & $\sigma(\PZ\PH)\times \BR(\PZ\to\Plp\Plm)$                & $\gHZZ^{2}$                   & $1.8\,\%$    & $1.3\,\%$     & \cite{Abramowicz:2016zbo} \\
    $\PZ\PH$            & $\sigma(\PZ\PH)\times \BR(\PZ\to\PQq\PAQq)$               & $\gHZZ^{2}$                   & $0.9\,\%$    & $0.6\,\%$     & \cite{Abramowicz:2016zbo} \\
    $\PZ\PH$            & $\sigma(\PZ\PH)\times \BR(\PH\to\PQb\PAQb)$               & $\gHZZ^{2}\gHbb^{2}/\GH$      & $0.41\,\%$   & $0.29\,\%$    & \cite{Abramowicz:2016zbo} \\
    $\PZ\PH$            & $\sigma(\PZ\PH)\times \BR(\PH\to\PQc\PAQc)$               & $\gHZZ^{2}\gHcc^2/\GH$        &  $7\,\%$    &  $5\,\%$      & \cite{Abramowicz:2016zbo} \\
    $\PZ\PH$            & $\sigma(\PZ\PH)\times \BR(\PH\to\Pg\Pg)$                  &                              &  $2.9\,\%$   & $2.1\,\%$     & \cite{Abramowicz:2016zbo} \\
    $\PZ\PH$            & $\sigma(\PZ\PH)\times \BR(\PH\to\tptm)$                   & $\gHZZ^{2}\gHTauTau^{2}/\GH$  &  $3.0\,\%$   & $2.1\,\%$     & \cite{Abramowicz:2016zbo} \\
    $\PZ\PH$            & $\sigma(\PZ\PH)\times \BR(\PH\to\PW\PW^*)$                & $\gHZZ^{2}\gHWW^{2}/\GH$      &  $2.4\,\%$   & $1.7\,\%$      & \cite{Abramowicz:2016zbo} \\
    $\PH\PGne\PAGne$    & $\sigma(\PH\PGne\PAGne)\times \BR(\PH\to\PQb\PAQb)$       & $\gHWW^{2}\gHbb^{2}/\GH$      &  $0.9\,\%$   & $0.6\,\%$     & \cite{Abramowicz:2016zbo} \\
    $\PH\PGne\PAGne$    & $\sigma(\PH\PGne\PAGne)\times \BR(\PH\to\PQc\PAQc)$       & $\gHWW^{2}\gHcc^{2}/\GH$      &  $12\,\%$     & $9\,\%$      & \cite{Abramowicz:2016zbo} \\
    $\PH\PGne\PAGne$    & $\sigma(\PH\PGne\PAGne)\times \BR(\PH\to\Pg\Pg)$          &                              &  $4.8\,\%$   &  $3.4\,\%$   & \cite{Abramowicz:2016zbo} \\    
    \bottomrule
  \end{tabular}
    \caption{Summary of the precisions obtainable for the Higgs
      observables with the first stage of CLIC for two integrated
      luminosity scenarios, corresponding to 50\,Hz running and 100\,Hz
      running, respectively; and assuming unpolarised beams.
      For the branching ratios, the measurement
      precision refers to the expected statistical uncertainty on the
      product of the relevant cross section and branching ratio; this
      is equivalent to the expected statistical uncertainty of the
      product of couplings divided by $\Gamma_{\PH}$ as indicated in
      the third column. \label{tab:GlobalFit:Input350}}
\end{table*}

\begin{table*}[htp]\centering
    \begin{tabular}{lllccc}\toprule
                        &                                                            &                             & \tabtt{Statistical precision} & \\ \cmidrule(l){4-5}
        \tabt{Channel}  & \tabt{Measurement}                                         & \tabt{Observable}           & $1.4\,\TeV$        & $3\,\TeV$ & Reference \\ 
                        &                                                            &                             & $4.0\,\abinv$      & $5.0\,\abinv$ & \\ \midrule
   $\PH\PGne\PAGne$     & $\PH\to\PQb\PAQb$ mass distribution                        & $\mH$                       &  $29\,\MeV$         & $28\,\MeV$ & \cite{Abramowicz:2016zbo} \\ \midrule        
   $\PZ\PH$             & $\sigma(\PZ\PH)\times \BR(\PH\to\PQb\PAQb)$                & $\gHZZ^{2}\gHbb^{2}/\GH$     & $2.0\,\%^{\dagger}$   & $4.3\,\%^{\dagger \ddagger}$ & \cite{Ellis:2017kfi} \\
   $\PH\PGne\PAGne$     & $\sigma(\PH\PGne\PAGne)\times \BR(\PH\to\PQb\PAQb)$        & $\gHWW^{2}\gHbb^{2}/\GH$     & $0.2\,\%$          & $0.2\,\%$ & \cite{Abramowicz:2016zbo} \\                 
   $\PH\PGne\PAGne$     & $\sigma(\PH\PGne\PAGne)\times \BR(\PH\to\PQc\PAQc)$        & $\gHWW^{2}\gHcc^{2}/\GH$     & $3.7\,\%$          & $4.4\,\%$ & \cite{Abramowicz:2016zbo} \\                 
   $\PH\PGne\PAGne$     & $\sigma(\PH\PGne\PAGne)\times \BR(\PH\to\Pg\Pg)$           &                             &  $3.1\,\%$          & $2.7\,\%$ & \cite{Abramowicz:2016zbo} \\                  
   $\PH\PGne\PAGne$     & $\sigma(\PH\PGne\PAGne)\times \BR(\PH\to\tptm)$            & $\gHWW^{2}\gHTauTau^{2}/\GH$ & $2.6\,\%$         & $2.8\,\%$ & \cite{Abramowicz:2016zbo} \\                 
   $\PH\PGne\PAGne$     & $\sigma(\PH\PGne\PAGne)\times \BR(\PH\to\mpmm)$            & $\gHWW^{2}\gHMuMu^{2}/\GH$   & $23\,\%$          & $16\,\%$ & \cite{Abramowicz:2016zbo} \\                  
   $\PH\PGne\PAGne$     & $\sigma(\PH\PGne\PAGne)\times \BR(\PH\to\upgamma\upgamma)$ &                             &  $9\,\%$          & $6\,\%^*$ & \cite{Abramowicz:2016zbo} \\                   
   $\PH\PGne\PAGne$     & $\sigma(\PH\PGne\PAGne)\times \BR(\PH\to\PZ\upgamma)$      &                             & $26\,\%$          & $19\,\%^*$ & \cite{Abramowicz:2016zbo} \\
   $\PH\PGne\PAGne$     & $\sigma(\PH\PGne\PAGne)\times \BR(\PH\to\PW\PW^*)$         & $\gHWW^{4}/\GH$             &  $0.6\,\%$         & $0.4\,\%^*$ & \cite{Abramowicz:2016zbo} \\                
   $\PH\PGne\PAGne$     & $\sigma(\PH\PGne\PAGne)\times \BR(\PH\to\PZ\PZ^*)$         & $\gHWW^{2}\gHZZ^{2}/\GH$     &  $3.4\,\%$         & $2.5\,\%^*$ & \cite{Abramowicz:2016zbo} \\               
   $\PH\epem$           & $\sigma(\PH\epem)\times \BR(\PH\to\PQb\PAQb)$              & $\gHZZ^{2}\gHbb^{2}/\GH$     &  $1.1\,\%$         & $1.5\,\%^*$ & \cite{Abramowicz:2016zbo} \\ \midrule      
   $\PQt\PAQt\PH$       & $\sigma(\PQt\PAQt\PH)\times \BR(\PH\to\PQb\PAQb)$          & $\gHtt^{2}\gHbb^{2}/\GH$     &  $4.5\,\%$         & $-$ & \cite{Abramowicz:2018rjq} \\                              
  \bottomrule
  \end{tabular}
    \caption{Summary of the precisions obtainable for the Higgs
    observables in the higher-energy CLIC stages with the updated 
    luminosities of $4.0\,\abinv$ at $\roots=1.4\,\TeV$, and
    $5.0\,\abinv$ at $\roots=3\,\TeV$. In both cases unpolarised beams
    have been assumed. 
    For $\gHtt$, the $3\,\TeV$ case has not yet been studied. 
    Numbers marked with $*$ are extrapolated from $\roots=1.4\,\TeV$
    to $\roots=3\,\TeV$ while $\dagger$ indicates projections based on fast simulations.
    For the branching ratios, the measurement precision refers to the expected
    statistical uncertainty on the product of the relevant cross
    section and branching ratio; this is equivalent to the expected
    statistical uncertainty of the product of couplings divided by
    $\Gamma_{\PH}$, as indicated in the third column. \label{tab:GlobalFit:Input143}
    $^{\ddagger}$ The value for $\sigma(\PZ\PH)\times \BR(\textrm{all hadronic})$ at 3\,TeV has been confirmed as 4\% in a full-simulation study \cite{Leogrande:2019dzm}.}
\end{table*}

\subsection{Combined fits}

Precisions on the Higgs couplings and width extracted from a model-independent
global fit, described in \cite{Abramowicz:2016zbo}, are given
in \autoref{tab:MIResultsPolarised8020} and \autoref{tab:MIResults100HzPolarised8020},
for 50\,Hz and 100\,Hz running at the initial stage, respectively.
The fit assumes the baseline scenario for beam polarisation,
i.e.\ equal amounts of --80\% and +80\% polarisation running throughout the initial energy stage,
and operation with $-80\,\%$ ($+80\,\%$) electron beam polarisation for 
$80\,\%$ ($20\,\%$) of the collected luminosity above 1\,\TeV. 
The increase in cross-section from having a predominantly negatively-polarised
electron beam is taken into account by multiplying the event rates for all
$\PW\PW$-fusion measurements by a factor of 1.48, corresponding to a factor of
1.8 for $80\,\%$ of the statistics and 0.2 for the remaining $20\,\%$. 
This approach is
conservative because it assumes that all backgrounds scale by the same amount; including those from
$s$-channel processes, which do not receive the same polarisation enhancement.

Each energy stage contributes significantly to the Higgs programme: 
the initial stage provides $\gHZZ$ and couplings to most fermions and
bosons, while the higher-energy stages improve them and add the top-quark,
muon, and photon couplings.  The precision on $\gHZZ$ is determined by the
statistics at the initial stage.

Precisions extracted from a model-dependent
global fit, also described in \cite{Abramowicz:2016zbo}, are given in \autoref{tab:MDResultsPolarised8020}
\autoref{tab:MDResults100HzPolarised8020} for 50\,Hz and 100\,Hz running at the initial stage, respectively.
This fit also assumes the baseline scenario for beam polarisation.

A global EFT fit has been carried out in \cite{deBlas:2019rxi} and updated in \cite{deBlas:2022ofj} for the purposes of comparing
future collider projects, and is extensively described there.
The corresponding projections for the increased CLIC integrated luminosities 
combined with the projected HL-LHC sensitivities, are given in \autoref{tab:eft-global1} and \autoref{tab:eft-global2} 
for the model SMEFT$_{\textrm{ND}}$, which does not assume flavour universality. 
The HL-LHC projections are also given separately for comparison.

\section{Conclusions}

Advancements in the CLIC accelerator design and a new baseline of 100\,Hz running at the initial stage result in significantly higher luminosity projections; this leads to an improvement on the Higgs coupling sensitivities of around a factor of two compared with the 2018 baseline, for measurements at the initial energy stage only, or at the first and second energy stages.

\appendix

\section*{Acknowledgements}

This work is done in the context of the CLICdp Collaboration, and rescales results from many
detailed analyses that are presented in \cite{Abramowicz:2016zbo}.

This work benefitted from services provided by the ILC Virtual Organisation, supported by the national resource providers of the EGI Federation. This research was done using resources provided by the Open Science Grid, which is supported by the National Science Foundation and the U.S. Department of Energy's Office of Science.

\clearpage

\begin{minipage}{\linewidth}
  \begin{minipage}{0.495\textwidth}
    \begin{table}[H]
\begin{tabular}{lccc}
\toprule
Parameter & \multicolumn{3}{c}{Relative precision}\\
\midrule
& $350\,\GeV$ & + $1.4\,\TeV$ & + $3\,\TeV$\\
&$2.2\,\abinv$& + $4\,\abinv$& + $5\,\abinv$\\
\midrule
$\gHZZ$ & 0.4\,\% & 0.4\,\% & 0.4\,\% \\
$\gHWW$ & 0.6\,\% & 0.4\,\% & 0.4\,\% \\
$\gHbb$ & 1.4\,\% & 0.5\,\% & 0.5\,\% \\
$\gHcc$ & 2.9\,\% & 1.4\,\% & 1.2\,\% \\
$\gHTauTau$ & 2.1\,\% & 1.0\,\% & 0.8\,\% \\
$\gHMuMu$ & $-$ & 9.6\,\% & 5.3\,\% \\
$\gHtt$ & $-$ & 2.3\,\% & 2.3\,\% \\
\midrule
$g^\dagger_{\PH\Pg\Pg}$ & 1.8\,\% & 1.0\,\% & 0.8\,\% \\
$g^\dagger_{\PH\PGg\PGg}$ & $-$ & 3.8\,\% & 2.2\,\% \\
$g^\dagger_{\PH\PZ\PGg}$ & $-$ & 10.6\,\% & 6.2\,\% \\
\midrule
$\Gamma_{\PH}$ & 3.2\,\% & 1.8\,\% & 1.7\,\% \\
\bottomrule
      \end{tabular}
      \caption{ }\label{tab:MIResultsPolarised8020}
    \end{table}
  \end{minipage}
  \begin{minipage}{0.495\textwidth}
    \begin{figure}[H]
      \includegraphics[width=\linewidth]{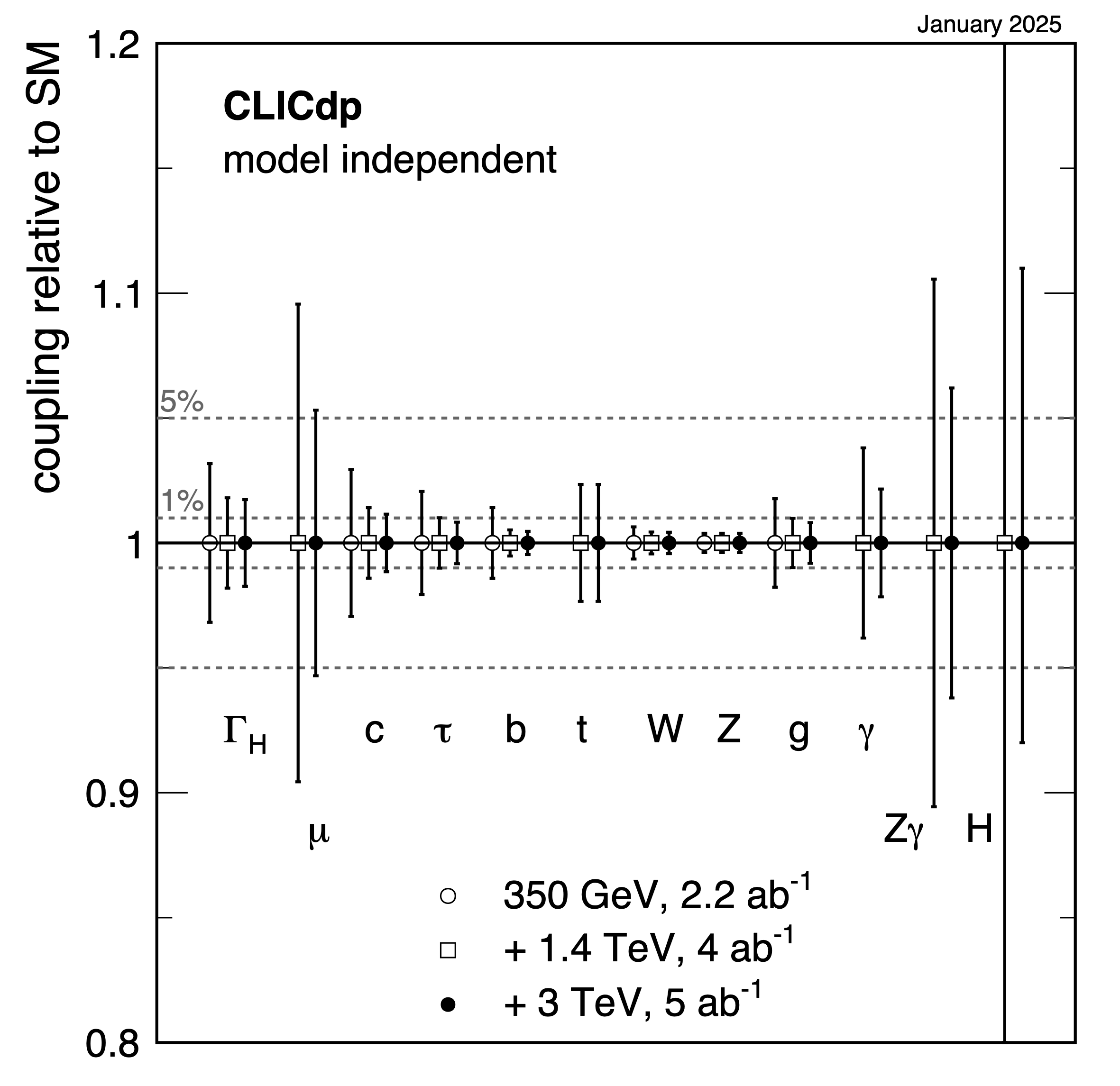}
      \caption{ }\label{fig:MIResultsPolarised8020}
    \end{figure}
  \end{minipage}
  Results of the model-independent fit assuming $2.2\,\abinv$ at $\sqrt{s}=350\,\GeV$,
  corresponding to 50\,Hz running.
        For $\gHtt$, the $3\,\TeV$ case has not yet been studied. The three
        effective couplings $g^\dagger_{\PH\Pg\Pg}$, 
        $g^\dagger_{\PH\PGg\PGg}$ and $g^\dagger_{\PH\PZ\PGg}$ are also included in the fit. 
        Operation with $-80\,\%$ ($+80\,\%$) electron beam polarisation is assumed for 
        $80\,\%$ ($20\,\%$) of the collected luminosity above 1\,\TeV, corresponding
        to the baseline scenario. 
\end{minipage}

\begin{minipage}{\linewidth}
  \begin{minipage}{0.495\textwidth}
    \begin{table}[H]
\begin{tabular}{lccc}
\toprule
Parameter & \multicolumn{3}{c}{Relative precision}\\
\midrule
& $350\,\GeV$ & + $1.4\,\TeV$ & + $3\,\TeV$\\
&$4.3\,\abinv$& + $4\,\abinv$& + $5\,\abinv$\\
\midrule
$\gHZZ$ & 0.3\,\% & 0.3\,\% & 0.3\,\% \\
$\gHWW$ & 0.5\,\% & 0.3\,\% & 0.3\,\% \\
$\gHbb$ & 1.0\,\% & 0.4\,\% & 0.4\,\% \\
$\gHcc$ & 2.1\,\% & 1.2\,\% & 1.0\,\% \\
$\gHTauTau$ & 1.5\,\% & 0.9\,\% & 0.7\,\% \\
$\gHMuMu$ & $-$ & 9.6\,\% & 5.3\,\% \\
$\gHtt$ & $-$ & 2.3\,\% & 2.3\,\% \\
\midrule
$g^\dagger_{\PH\Pg\Pg}$ & 1.3\,\% & 0.8\,\% & 0.7\,\% \\
$g^\dagger_{\PH\PGg\PGg}$ & $-$ & 3.8\,\% & 2.2\,\% \\
$g^\dagger_{\PH\PZ\PGg}$ & $-$ & 10.6\,\% & 6.2\,\% \\
\midrule
$\Gamma_{\PH}$ & 2.3\,\% & 1.4\,\% & 1.3\,\% \\
\bottomrule
      \end{tabular}
      \caption{ }\label{tab:MIResults100HzPolarised8020}
    \end{table}
  \end{minipage}
  \begin{minipage}{0.495\textwidth}
    \begin{figure}[H]
      \includegraphics[width=\linewidth]{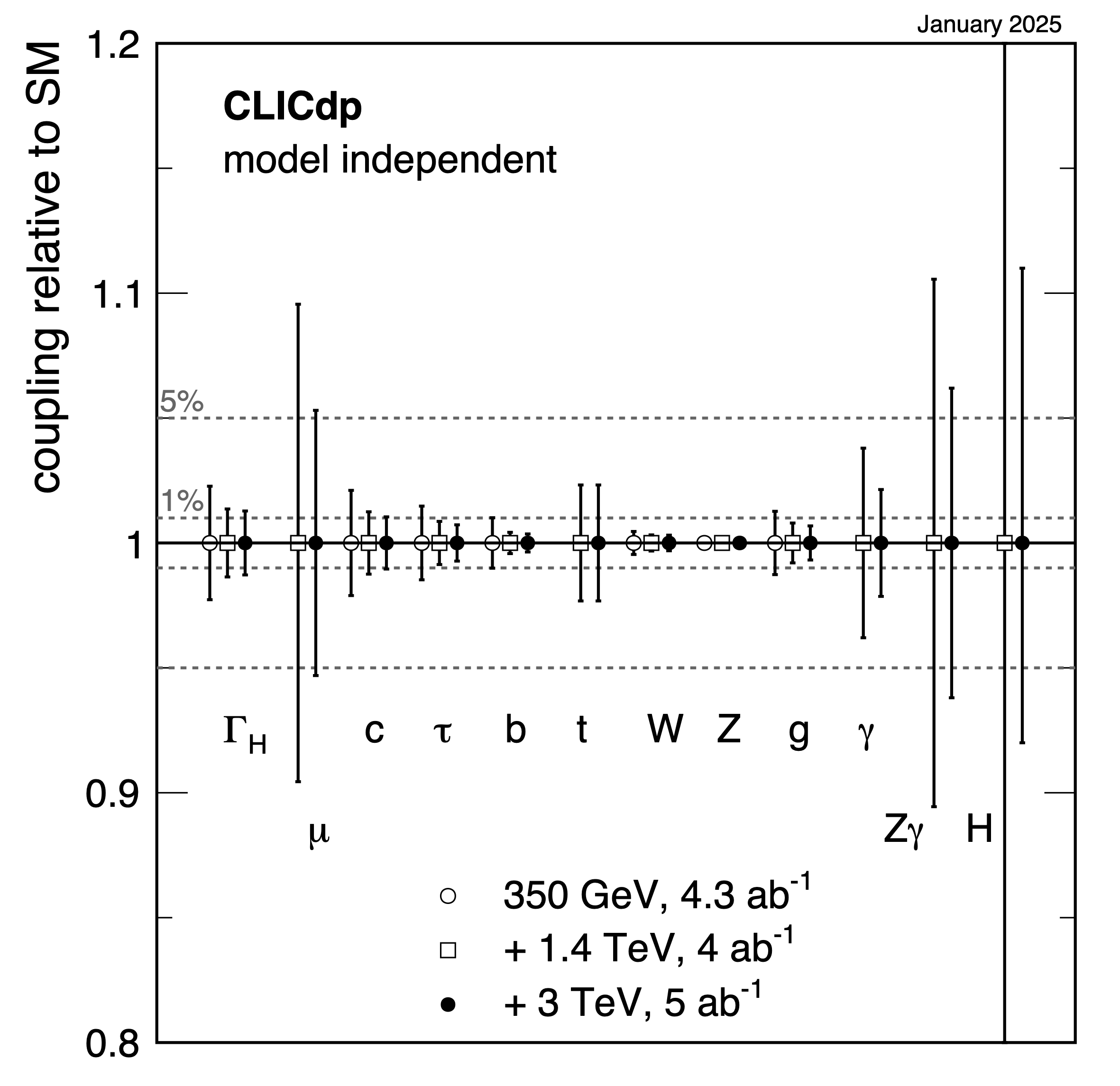}
      \caption{ }\label{fig:MIResults100HzPolarised8020}
    \end{figure}
  \end{minipage}
  Results of the model-independent fit assuming $4.3\,\abinv$ at $\sqrt{s}=350\,\GeV$,
  corresponding to 100\,Hz running (the updated CLIC baseline).
        For $\gHtt$, the $3\,\TeV$ case has not yet been studied. The three
        effective couplings $g^\dagger_{\PH\Pg\Pg}$, 
        $g^\dagger_{\PH\PGg\PGg}$ and $g^\dagger_{\PH\PZ\PGg}$ are also included in the fit. 
        Operation with $-80\,\%$ ($+80\,\%$) electron beam polarisation is assumed for 
        $80\,\%$ ($20\,\%$) of the collected luminosity above 1\,\TeV, corresponding
        to the baseline scenario. 
\end{minipage}

\begin{minipage}{\linewidth}
\vspace*{1cm}
  \begin{minipage}{0.495\textwidth}
    \begin{table}[H]
\begin{tabular}{lccc}
\toprule
Parameter & \multicolumn{3}{c}{Relative precision}\\
\midrule
& $350\,\GeV$ & + $1.4\,\TeV$& + $3\,\TeV$\\
&$2.2\,\abinv$& + $4\,\abinv$& + $5\,\abinv$\\
\midrule
$\kappa_{\PH\PZ\PZ}$ & 0.3\,\% & 0.2\,\% & 0.2\,\% \\
$\kappa_{\PH\PW\PW}$ & 0.5\,\% & 0.1\,\% & 0.1\,\% \\
$\kappa_{\PH\PQb\PQb}$ & 0.9\,\% & 0.2\,\% & 0.2\,\% \\
$\kappa_{\PH\PQc\PQc}$ & 2.7\,\% & 1.3\,\% & 1.1\,\% \\
$\kappa_{\PH\PGt\PGt}$ & 1.9\,\% & 0.9\,\% & 0.7\,\% \\
$\kappa_{\PH\PGm\PGm}$ & $-$ & 9.6\,\% & 5.3\,\% \\
$\kappa_{\PH\PQt\PQt}$ & $-$ & 2.3\,\% & 2.3\,\% \\
$\kappa_{\PH\Pg\Pg}$ & 1.4\,\% & 0.9\,\% & 0.7\,\% \\
$\kappa_{\PH\PGg\PGg}$ & $-$ & 3.8\,\% & 2.1\,\% \\
$\kappa_{\PH\PZ\PGg}$ & $-$ & 10.6\,\% & 6.2\,\% \\
\bottomrule
\end{tabular}
\caption{ }\label{tab:MDResultsPolarised8020}
    \end{table}
  \end{minipage}
  \begin{minipage}{0.495\textwidth}
   \begin{figure}[H]
     \includegraphics[width=\linewidth]{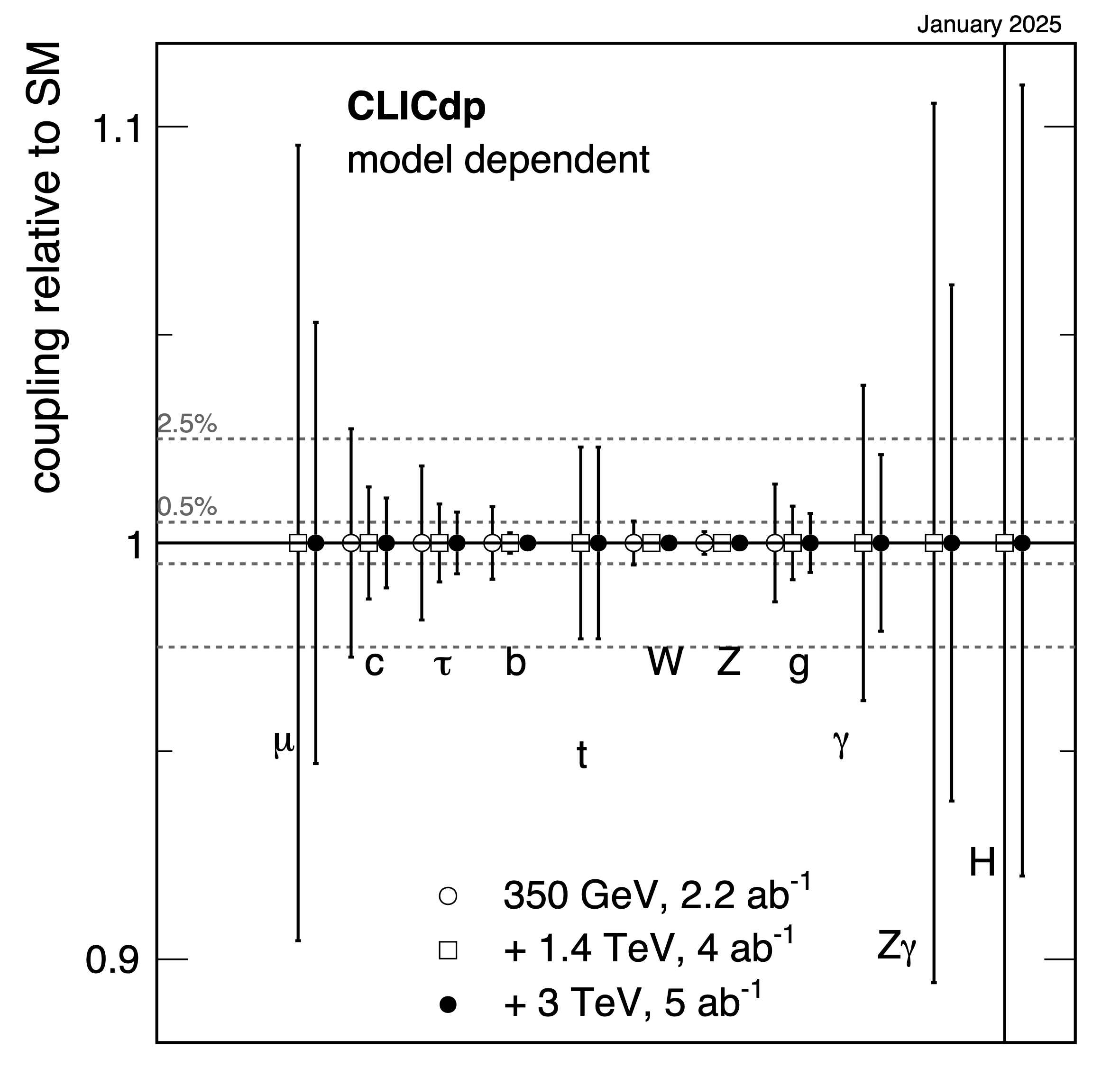}
     \caption{ }\label{fig:MDResultsPolarised8020}
   \end{figure}
  \end{minipage}
  Results of the model-dependent fit without theoretical uncertainties, assuming $2.2\,\abinv$ at $\sqrt{s}=350\,\GeV$, 
  corresponding to 50\,Hz running.
  For $\kappa_{\PH\PQt\PQt}$, the $3\,\TeV$ case has not yet been
  studied. The uncertainty of the total width is calculated from the fit results. 
  Operation with $-80\,\%$ ($+80\,\%$) electron beam polarisation is assumed for 
  $80\,\%$ ($20\,\%$) of the collected luminosity above 1\,\TeV, corresponding to the baseline scenario. 
\end{minipage}

\begin{minipage}{\linewidth}
\vspace*{1cm}
  \begin{minipage}{0.495\textwidth}
    \begin{table}[H]
\begin{tabular}{lccc}
\toprule
Parameter & \multicolumn{3}{c}{Relative precision}\\
\midrule
& $350\,\GeV$ & + $1.4\,\TeV$& + $3\,\TeV$\\
&$4.3\,\abinv$& + $4\,\abinv$& + $5\,\abinv$\\
\midrule
$\kappa_{\PH\PZ\PZ}$ & 0.2\,\% & 0.1\,\% & 0.1\,\% \\
$\kappa_{\PH\PW\PW}$ & 0.4\,\% & 0.1\,\% & 0.1\,\% \\
$\kappa_{\PH\PQb\PQb}$ & 0.6\,\% & 0.2\,\% & 0.2\,\% \\
$\kappa_{\PH\PQc\PQc}$ & 2.0\,\% & 1.2\,\% & 1.0\,\% \\
$\kappa_{\PH\PGt\PGt}$ & 1.3\,\% & 0.8\,\% & 0.7\,\% \\
$\kappa_{\PH\PGm\PGm}$ & $-$ & 9.6\,\% & 5.3\,\% \\
$\kappa_{\PH\PQt\PQt}$ & $-$ & 2.3\,\% & 2.3\,\% \\
$\kappa_{\PH\Pg\Pg}$ & 1.0\,\% & 0.7\,\% & 0.6\,\% \\
$\kappa_{\PH\PGg\PGg}$ & $-$ & 3.8\,\% & 2.1\,\% \\
$\kappa_{\PH\PZ\PGg}$ & $-$ & 10.6\,\% & 6.2\,\% \\
\bottomrule
\end{tabular}
\caption{ }\label{tab:MDResults100HzPolarised8020}
    \end{table}
  \end{minipage}
  \begin{minipage}{0.495\textwidth}
   \begin{figure}[H]
     \includegraphics[width=\linewidth]{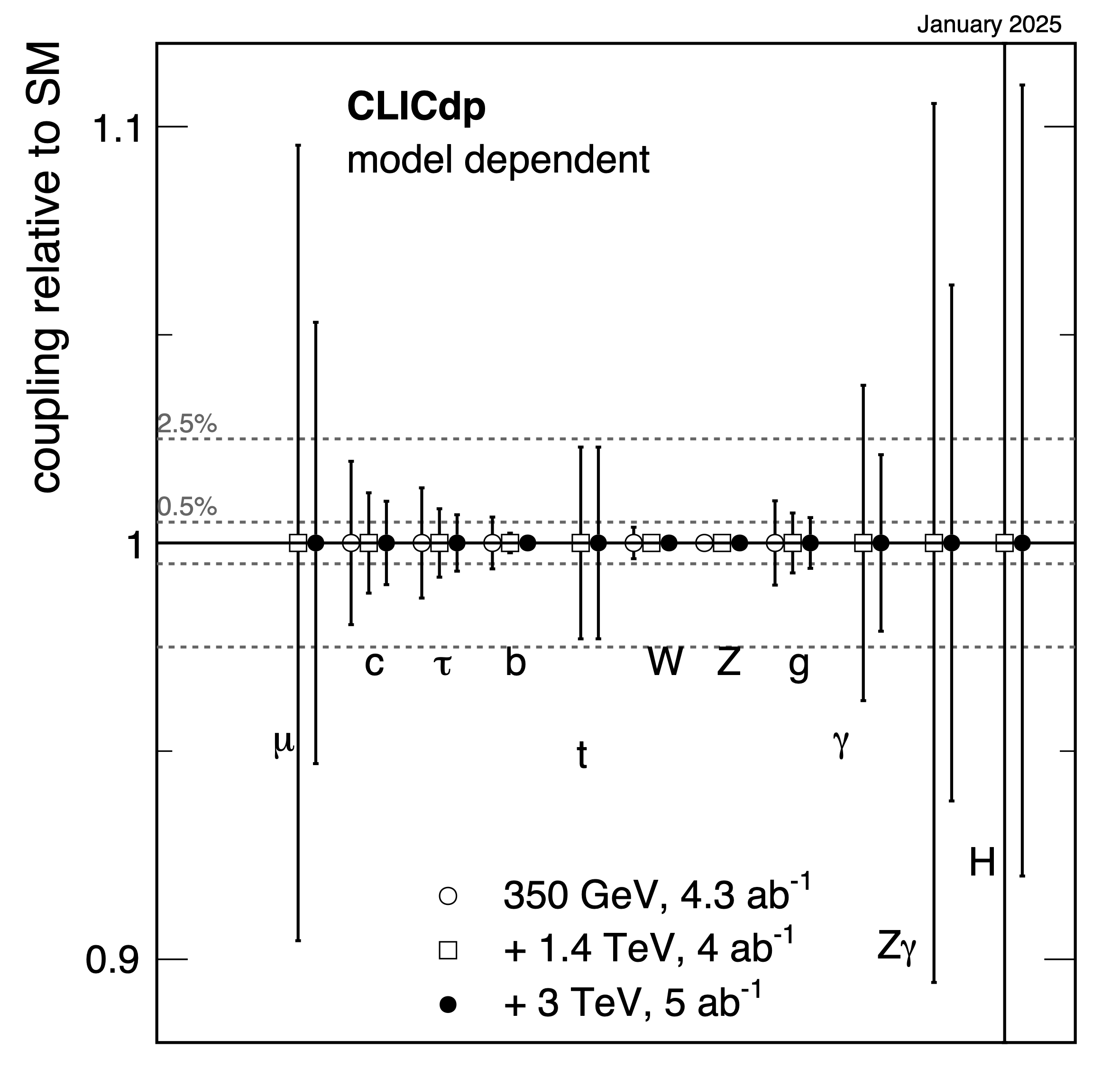}
     \caption{ }\label{fig:MDResults100HzPolarised8020}
   \end{figure}
  \end{minipage}
  Results of the model-dependent fit without theoretical uncertainties, assuming $4.3\,\abinv$ at $\sqrt{s}=350\,\GeV$, 
  corresponding to 100\,Hz running (the updated CLIC baseline).
  For $\kappa_{\PH\PQt\PQt}$, the $3\,\TeV$ case has not yet been
  studied. The uncertainty of the total width is calculated from the fit results. 
  Operation with $-80\,\%$ ($+80\,\%$) electron beam polarisation is assumed for 
  $80\,\%$ ($20\,\%$) of the collected luminosity above 1\,\TeV, corresponding to the baseline scenario. 
\end{minipage}

\begin{table}[t]
\centering
{ 
\begin{tabular}{ c | c | c | c c c }
\toprule
   &Benchmark   &HL-LHC   &\multicolumn{3}{c}{HL-LHC + CLIC} \\
   &    & & 380\,GeV   &  + 1.5\,TeV  & + 3\,TeV  \\
   &    & & 2.2\, ab$^{-1}$   &  4\, ab$^{-1}$  & 5\, ab$^{-1}$  \\
\midrule
 \hspace{-0.2cm}$\gHZZ^{\mathrm{eff}}[\%]$& SMEFT$_{\textrm{ND}}$    &  2.2   & 0.36     & 0.15     &  0.13   \\
 \hspace{-0.2cm}$\gHWW^{\mathrm{eff}}[\%]$& SMEFT$_{\textrm{ND}}$    &  2.0   & 0.36    & 0.15    &  0.13   \\
 \hspace{-0.2cm}$g_{\PH\PGg\PGg}^{\mathrm{eff}}[\%]$& SMEFT$_{\textrm{ND}}$    &  2.5   &  1.2   & 1.1    &  1.0   \\
 \hspace{-0.2cm}$g_{\PH\PZ\PGg}^{\mathrm{eff}}[\%]$& SMEFT$_{\textrm{ND}}$   &  11  &  10   &  3.8   &  3.2   \\
 \hspace{-0.2cm}$g_{\PH\Pg\Pg}^{\mathrm{eff}}[\%]$& SMEFT$_{\textrm{ND}}$   &  1.8   &  0.86   & 0.62    & 0.54    \\
 \hspace{-0.2cm}$\gHtt^{\mathrm{eff}}[\%]$& SMEFT$_{\textrm{ND}}$   &  3.5*   &  --   & 1.9*   &  1.9*   \\
 \hspace{-0.2cm}$\gHcc^{\mathrm{eff}}[\%]$& SMEFT$_{\textrm{ND}}$   &  --   &  2.7   &  1.4   &  1.2   \\
 \hspace{-0.2cm}$\gHbb^{\mathrm{eff}}[\%]$& SMEFT$_{\textrm{ND}}$   &  4.5   &   0.77  &  0.39   &  0.35   \\
 \hspace{-0.2cm}$\gHTauTau^{\mathrm{eff}}[\%]$& SMEFT$_{\textrm{ND}}$   &  2.3   &  1.2   & 0.81    &  0.68   \\
 \hspace{-0.2cm}$\gHMuMu^{\mathrm{eff}}[\%]$& SMEFT$_{\textrm{ND}}$   &   5.6  &  5.2   & 4.6    &  3.8   \\
\midrule
 \hspace{-0.2cm}$\delta g_{1\PZ}[\times 10^{2}]$& SMEFT$_{\textrm{ND}}$  &    0.31 &  0.069   & 0.027    &  0.024   \\
 \hspace{-0.2cm}$\delta \kappa_{ \PGg}[\times 10^{2}]$& SMEFT$_{\textrm{ND}}$  & 0.97    &  0.1   &  0.038   &  0.032   \\
 \hspace{-0.2cm}$\lambda_{\PZ}[\times 10^{2}]$& SMEFT$_{\textrm{ND}}$ &   0.4  &   0.017  & 0.002    &  0.0009   \\
\bottomrule
\end{tabular}
}
\caption{\label{tab:eft-global1}  
Sensitivity at 68\% probability to deviations in the different effective Higgs couplings and anomalous triple gauge couplings from a global SMEFT fit, assuming $2.2\,\abinv$ at $\sqrt{s}=350\,\GeV$, 
  corresponding to 50\,Hz running, using the benchmark SMEFT$_{\textrm{ND}}$ described in \cite{deBlas:2019rxi} and updated in \cite{deBlas:2022ofj}. (The information about the other degrees of freedom included in the SMEFT$_{\textrm{ND}}$ fit in \cite{deBlas:2019rxi}, i.e. $g_{L,R}^f$, is omitted in this table.) These numbers can be compared to those of Table\,29 in \cite{deBlas:2022ofj}. As in \cite{deBlas:2022ofj}, for the CLIC columns we include the results from the Z boson radiative return events. For $\epem \to \PWp\PWm$, we also include the statistical optimal observable results discussed in ~\cite{DeBlas:2019qco,deBlas:2022ofj}, updated to the luminosities discussed in this note.  
* For the top Yukawa coupling, coming from the $\PQt\PQt\PH$ process where additional EFT interactions not considered in SMEFT$_{\textrm{ND}}$ contribute, we keep the more conservative results in \cite{deBlas:2019rxi} and combine them with the CLIC precision.
}
\end{table}

\begin{table}[t]
\centering
{ 
\begin{tabular}{ c | c | c | c c c }
\toprule
   &Benchmark   &HL-LHC   &\multicolumn{3}{c}{HL-LHC + CLIC} \\
   &    & & 380\,GeV   &  + 1.5\,TeV  & + 3\,TeV  \\
   &    & & 4.3\, ab$^{-1}$   &  4\, ab$^{-1}$  & 5\, ab$^{-1}$  \\
\midrule
 \hspace{-0.2cm}$\gHZZ^{\mathrm{eff}}[\%]$& SMEFT$_{\textrm{ND}}$    &  2.2   &  0.28   & 0.14    &  0.13   \\
 \hspace{-0.2cm}$\gHWW^{\mathrm{eff}}[\%]$& SMEFT$_{\textrm{ND}}$    &  2.0   &  0.28   &  0.14   &   0.13  \\
 \hspace{-0.2cm}$g_{\PH\PGg\PGg}^{\mathrm{eff}}[\%]$& SMEFT$_{\textrm{ND}}$    &  2.5   &  1.2   &  1.1   &  0.99   \\
 \hspace{-0.2cm}$g_{\PH\PZ\PGg}^{\mathrm{eff}}[\%]$& SMEFT$_{\textrm{ND}}$   &  11   &  9.7   & 3.1    &  2.6   \\
 \hspace{-0.2cm}$g_{\PH\Pg\Pg}^{\mathrm{eff}}[\%]$& SMEFT$_{\textrm{ND}}$   &   1.8  & 0.73    &  0.56   &  0.5   \\
 \hspace{-0.2cm}$\gHtt^{\mathrm{eff}}[\%]$& SMEFT$_{\textrm{ND}}$   &  3.5*   &  --   & 1.9*   &  1.9*   \\
 \hspace{-0.2cm}$\gHcc^{\mathrm{eff}}[\%]$& SMEFT$_{\textrm{ND}}$   &  --   &   2.0  & 1.3    &  1.1   \\
 \hspace{-0.2cm}$\gHbb^{\mathrm{eff}}[\%]$& SMEFT$_{\textrm{ND}}$   &   4.5  &  0.62   &  0.39   & 0.35    \\
 \hspace{-0.2cm}$\gHTauTau^{\mathrm{eff}}[\%]$& SMEFT$_{\textrm{ND}}$   &  2.3   &  1.0   & 0.73    &  0.63   \\
 \hspace{-0.2cm}$\gHMuMu^{\mathrm{eff}}[\%]$& SMEFT$_{\textrm{ND}}$   &  5.6   &   5.2  &  4.6   &  3.8   \\
\midrule
 \hspace{-0.2cm}$\delta g_{1\PZ}[\times 10^{2}]$& SMEFT$_{\textrm{ND}}$  &   0.31  &  0.060   & 0.023    &  0.021   \\
 \hspace{-0.2cm}$\delta \kappa_{ \PGg}[\times 10^{2}]$& SMEFT$_{\textrm{ND}}$  &  0.97   &   0.084  &  0.033   &  0.027   \\
 \hspace{-0.2cm}$\lambda_{\PZ}[\times 10^{2}]$& SMEFT$_{\textrm{ND}}$ &  0.4   &   0.012  &  0.002   &  0.0009   \\
\bottomrule
\end{tabular}
}
\caption{\label{tab:eft-global2} 
Sensitivity at 68\% probability to deviations in the different effective Higgs couplings and anomalous triple gauge couplings from a global SMEFT fit, assuming $4.3\,\abinv$ at $\sqrt{s}=350\,\GeV$, 
  corresponding to 100\,Hz running (the updated CLIC baseline), using the benchmark SMEFT$_{\textrm{ND}}$ described in \cite{deBlas:2019rxi} and updated in \cite{deBlas:2022ofj}. (The information about the other degrees of freedom included in the SMEFT$_{\textrm{ND}}$ fit in \cite{deBlas:2019rxi}, i.e. $g_{L,R}^f$, is omitted in this table.) These numbers can be compared to those of Table\,29 in \cite{deBlas:2022ofj}. As in \cite{deBlas:2022ofj}, for the CLIC columns we include the results from the Z boson radiative return events. For $\epem \to \PWp\PWm$, we also include the statistical optimal observable results discussed in ~\cite{DeBlas:2019qco,deBlas:2022ofj}, updated to the luminosities discussed in this note.  
* For the top Yukawa coupling, coming from the $\PQt\PQt\PH$ process where additional EFT interactions not considered in SMEFT$_{\textrm{ND}}$ contribute, we keep the more conservative results in \cite{deBlas:2019rxi} and combine them with the CLIC precision.
}
\end{table}


\printbibliography[title=References]

\end{document}